\documentclass[12pt]{article}
\usepackage{amsmath,color} 
\input amssym

\def\red#1{{\color{red} #1}}

\begin{document}

\def\prg#1{\par\medskip\noindent{\bf #1}}  \def\ra{\rightarrow}
\newcounter{nbr}
\def\note#1{\bitem\vspace{-5pt}\addtocounter{nbr}{1}
            \item{} #1\vspace{-5pt}
            \eitem}
\def\lra{\leftrightarrow}           \def\Ra{\Rightarrow}
\def\nin{\noindent}                 \def\pd{\partial}
\def\dis{\displaystyle}             \def\Lra{{\Leftrightarrow}}
\def\grl{{GR$_\Lambda$}}            \def\tmgl{\hbox{TMG$_\Lambda$}}
\def\vsm{\vspace{-9pt}}             \def\vsmb{\vspace{-5pt}}
\def\cs{{\scriptstyle\rm CS}}       \def\ads3{{\rm AdS$_3$}}
\def\Leff{\hbox{$\mit\L_{\hspace{.6pt}\rm eff}\,$}}
\def\bull{\raise.25ex\hbox{\vrule height.8ex width.8ex}}
\def\Lie{{\cal L}\hspace{-.7em}\raise.25ex\hbox{--}\hspace{.2em}}
\def\sS{\hspace{2pt}S\hspace{-0.83em}\diagup}
\def\ric{{Ric}}
\def\nb{ \marginpar{\bf\Large ?} }  \def\hd{{^\star}}
\def\dis{\displaystyle}             \def\mb#1{\hbox{{\boldmath $#1$}}}
\def\ul#1{\underline{#1}}           \def\ub#1{\underbrace{#1}}
\def\phb{\phantom{\Big|}}           \def\gr3{\hbox{GR${}_3$}}
\def\chm{\checkmark}                \def\chmr{\red{\chm}}
\def\ir#1{\,{}^{#1}\hspace{-1.2pt}}
\def\irr#1{\,{}^{(#1)}\hspace{-1.2pt}}

\def\hook{\hbox{\vrule height0pt width4pt depth0.3pt
\vrule height7pt width0.3pt depth0.3pt
\vrule height0pt width2pt depth0pt}\hspace{0.8pt}}
\def\inn{\hook}
\def\first{\rm (1ST)}  \def\second{\hspace{-1cm}\rm (2ND)}

\def\G{\Gamma}        \def\S{\Sigma}        \def\L{{\mit\Lambda}}
\def\D{\Delta}        \def\Th{\Theta}
\def\a{\alpha}        \def\b{\beta}         \def\g{\gamma}
\def\d{\delta}        \def\m{\mu}           \def\n{\nu}
\def\th{\theta}       \def\k{\kappa}        \def\l{\lambda}
\def\vphi{\varphi}    \def\ve{\varepsilon}  \def\p{\pi}
\def\r{\rho}          \def\Om{\Omega}       \def\om{\omega}
\def\s{\sigma}        \def\t{\tau}          \def\eps{\epsilon}
\def\nab{\nabla}      \def\btz{{\rm BTZ}}   \def\heps{{\hat\eps}}

\def\bT{\bar{T}}      \def\hH{\widehat{H}}  \def\hE{\widehat{E}}
\def\tG{{\tilde G}}   \def\cF{{\cal F}}    \def\cA{{\cal A}}
\def\cL{{\cal L}}     \def\cM{{\cal M }}   \def\cE{{\cal E}}
\def\cH{{\cal H}}     \def\hcH{\hat{\cH}}  \def\cT{{\cal T}}
\def\hA{\hat{A}}      \def\hB{\hat{B}}     \def\hK{\hat{K}}
\def\cK{{\cal K}}     \def\hcK{\hat{\cK}}  \def\cT{{\cal T}}
\def\cO{{\cal O}}     \def\hcO{\hat{\cal O}} \def\cV{{\cal V}}
\def\tom{{\tilde\omega}}  \def\cE{{\cal E}} \def\bH{\bar{H}}
\def\cR{{\cal R}}    \def\hR{{\hat R}{}}   \def\hL{{\hat\L}}
\def\tb{{\tilde b}}  \def\tA{{\tilde A}}   \def\hom{{\hat\om}}
\def\tT{{\tilde T}}  \def\tR{{\tilde R}}   \def\tcL{{\tilde\cL}}
\def\he{{\hat e}}    \def\hom{{\hat\om}}   \def\hth{\hat\theta}
\def\hxi{\hat\xi}    \def\hg{\hat g}       \def\hb{{\hat b}}
\def\tH{{\tilde H}}  \def\tV{{\tilde V}}   \def\ha{{\bar a}}
\def\hb{{\bar b}}    \def\bR{\bar{R}}      \def\bF{{\bar F}}
\def\haa{{\bar\a}}   \def\hbb{{\bar\b}}    \def\hgg{{\bar\g}}
\def\tPhi{{\tilde\Phi}} \def\barb{{\bar b}} \def\tPsi{{\tilde\Psi}}
\def\bK{{\bar K}}    \def\bk{{\bar k}}     \def\orth{{\perp}}
\def\bi{{\bar\imath}} \def\bj{{\bar\jmath}} \def \bk{{\bar k}}
\def\bm{{\bar m}}     \def\bn{{\bar n}}    \def\bl{{\bar l}}
\let\Pi\varPi         \def\chH{\check{H}}  \def\bB{{\bar B}}
\let\eR\varOmega    \def\hpi{{\hat\pi}}    \def\hPi{{\hat\Pi}}
\def\cN{{\cal N}}    \def\bs{{\bar s}}     \def\hN{{\hat N}}
\def\hH{{\hat H}}    \def\hb{{\hat b}}     \def\hom{{\hat\omega}}

\vfuzz=2pt 
\def\nn{\nonumber}
\def\be{\begin{equation}}             \def\ee{\end{equation}}
\def\ba#1{\begin{array}{#1}}          \def\ea{\end{array}}
\def\bea{\begin{eqnarray} }           \def\eea{\end{eqnarray} }
\def\beann{\begin{eqnarray*} }        \def\eeann{\end{eqnarray*} }
\def\beal{\begin{eqalign}}            \def\eeal{\end{eqalign}}
\def\lab#1{\label{eq:#1}}             \def\eq#1{(\ref{eq:#1})}
\def\bsubeq{\begin{subequations}}     \def\esubeq{\end{subequations}}
\def\bitem{\begin{itemize}}           \def\eitem{\end{itemize}}
\renewcommand{\theequation}{\thesection.\arabic{equation}}

\title{Entropy in three-dimensional general relativity: Kerr-AdS black hole}

\author{M. Blagojevi\'c and B. Cvetkovi\'c\footnote{
        Email addresses: \texttt{mb@ipb.ac.rs, cbranislav@ipb.ac.rs}} \\
Institute of Physics, University of Belgrade,\\
                      Pregrevica 118, 11080 Belgrade-Zemun, Serbia}
\date{\today}
\maketitle

\begin{abstract}
Black hole thermodynamics of the Kerr-AdS spacetime in three-dimensional general relativity is analyzed using a Hamiltonian approach.
The values of the conserved charges and entropy, obtained by a proper treatment of the AdS asymptotic conditions, are shown to satisfy the first law of black hole dynamics.

\end{abstract}
\section{Introduction}
\setcounter{equation}{1}

Asymptotic conditions play a crucial role in understanding basic features of the black hole thermodynamics. In the case of asymptotically flat black holes in general relativity (GR), the first law is shown to hold for any stationary and axially symmetric black hole with a bifurcate Killing horizon \cite{x1}. The situation with black holes in an asymptotically anti-de Sitter (AdS) background is more involved. As discussed by Gibbons et al. \cite{x2} for Kerr-AdS black holes, the relation between conserved charges, entropy and the angular velocity on one side, and the first law on the other, is not fully settled.

In order to clearly understand thermodynamics of Kerr-AdS black holes, one is naturally led to consider the corresponding three-dimensional (3D) models, as they allow us to investigate the problem in a technically much simpler context. A detailed study of the subject can be found in Hawking et al. \cite{x3}, where the authors analyzed, inter alia, how different coordinate systems affect the form of thermodynamic variables; see also \cite{x4,x5}. However, some aspects of their analysis of energy and the first law require further consideration.

A Hamiltonian approach to  black hole entropy has been recently proposed in Ref. \cite{x6}, and applied to asymptotically flat Kerr black holes in Ref. \cite{x7}. In the present paper, we use the same approach to analyze thermodynamic properties of the Kerr-AdS black hole in three-dimensional GR (\gr3). Our analysis shows that a proper treatment of the AdS asymptotic conditions ensures a consistency between the conserved charges, angular velocity and black hole entropy, expressed by the validity of the first law.

\section{Hamiltonian approach to entropy}
\setcounter{equation}{0}

Although the Hamiltonian approach to entropy \cite{x6,x7} is focused on black holes in Poincar\'e gauge theory \cite{x8}, where both the torsion and the curvature define the gravitational dynamics, it can be equally well used in the realm of GR as a Riemannian theory of gravity.

As a preparation for such an approach in 3D spacetime, we introduce the first-order orthonormal frame formulation of \gr3, in which the coframe $b^i=b^i{_\m}dx^\m$ and the antisymmetric spin connection $\om^{ij}=\om^{ij}{_\m}dx^\m$ (1-forms) are \emph{independent} dynamical variables. The related field strengths are the torsion $T^i=d b^i +\om^{ij}{_k}b^k\equiv\nab b^i$ and the curvature $R^{ij}=d\om^{ij}+\om^i{_k}\om^{kj}$ (2-forms), and metric is defined
by $g=\eta_{ij}b^i\otimes b^j$, with $\eta_{ij}=(1,-1,-1)$.
In the absence of matter, the gravitational dynamics is defined by the  Lagrangian 3-form
\be
L_G=-a_0\ve_{ijk}b^i R^{jk}-\frac{1}{3}\L_0\ve_{ijk}b^ib^jb^k\, ,
\ee
where $a_0=1/16\pi$ (in units $G=1$), $\L_0$ is a cosmological constant, and the completely antisymmetric tensor $\ve_{ijk}$ is normalized by $\ve_{012}=1$.
Varying $L_G$ with respect to $b^i$ and $\om^{ij}$, one obtains the field equations of \gr3\ in vacuum,
\bea
\d b^i:&&-a_0\ve_{ijk}R^{jk}-\L_0\ve_{ijk}b^jb^k=0\,,                 \nn\\
\d\om^{ij}&&-2a_0\ve_{ijk}\nab b^k=0\, .
\eea
The second equation yields the condition for vanishing torsion, $\nab b^k=0$, whereupon the first equation takes the standard form of Einstein equation with a cosmological constant.

Our approach to black hole entropy is an extension of the Hamiltonian treatment of conserved charges as boundary terms at infinity. In the case of a black hole solution, we assume that the boundary of the spatial section of spacetime $\S$ has \emph{two components}, one at infinity and the other at horizon, $\pd\S=S_\infty\cup S_H$. As a consequence, the boundary term $\G$ has two parts, $\G:=\G_\infty-\G_H$ (the minus sign in front of $\G_H$ reflects the change in orientation). For stationary and axially symmetric black holes in \gr3, with Killing vectors $\pd_t$ and $\pd_\vphi$, these boundary terms can be defined by the \emph{variational equations} \cite{x6,x7}
\bsubeq\lab{2.3}
\bea
\d\G_\infty&=&\oint_{S_\infty}\d B(\xi)\,,\qquad
       \d\G_H=\oint_{S_H} \d B(\xi)\,,                                  \\
\d B(\xi)&:=&\frac{1}{2}(\xi\inn\om^{ij})\d H_{ij}
   +\frac{1}{2}\d\om^{ij}(\xi\inn H_{ij})\, ,
\eea
where $\xi$ is a Killing vector on $S_\infty$ or $S_H$, and $H_{ij}$ is the covariant momentum
\be
H_{ij}:=\frac{\pd L_G}{\pd R^{ij}}=-2a_0\ve_{ijk}b^k\,.
\ee
As a consequence,
\be
\d B(\xi)=-a_0\ve_{ijk}\Big[(\xi\inn\om^{ij})\d b^k
                              +\d\om^{ij}(\xi\inn b^k)\Big]\,.
\ee
\esubeq

The variational procedure is assumed to satisfy the following rules:
\bitem\vsm
\item[(a1)] the variation $\d\G_\infty$ is performed over a suitable set of asymptotic states, leaving the background configuration fixed;\vsm
\item[(a2)] the variation $\d\G_H$ is performed by varying parameters of a solution, but keeping surface gravity constant (the zeroth law of black hole thermodynamics).
\eitem
If, under the adopted boundary conditions, $\G_\infty$ and $\G_H$ are \emph{finite solutions} of the variational equations \eq{2.3}, $\G_\infty$ is interpreted as the \emph{asymptotic charge} (energy-momentum or angular momentum) whereas $\G_H$ defines \emph{entropy} as the canonical charge on horizon.  However, if Eqs. \eq{2.3} are not integrable or their solutions are divergent, the boundary conditions have to be reconsidered. The validity of the first law follows from the differentiability of the canonical generator.

\section{Geometry of Kerr-AdS spacetime}
\setcounter{equation}{0}

Here, we describe geometry of the Kerr-AdS black hole in two different coordinate systems.

\prg{Part 1.} In 3D spacetime, the metric of the Kerr-AdS black can be described in the Boyer-Lindquist coordinates $(t,r,\vphi)$ \cite{x3} as
\bsubeq\lab{3.1}
\bea
ds^2&=&\frac{\D}{r^2}\Big(dt+\frac{a}{\a}d\vphi\Big)^2-\frac{r^2}{\D}dr^2
      -\frac{1}{r^2}\Big[adt+\frac{(r^2+a^2)}{\a}d\vphi\Big]^2       \nn\\
 &=&\frac{\D-a^2}{r^2}dt^2-\frac{2a}{r^2\a}\big(r^2+a^2-\D\big)dt d\vphi
    -\frac{\S^2}{r^2\a^2}d\vphi^2-\frac{r^2}{\D}dr^2\,,
\eea
where
\be
\D=(r^2+a^2)(1+\l r^2)-2mr^2\,,\qquad\S^2=(r^2+a^2)^2-a^2\D\,,\qquad
\a:=1-\l a^2\, ,
\ee
\esubeq
This metric is similar to its 4D counterpart, and it solves the field equations for $a_0\l+\L_0=0$. The event horizon is defined by the larger root of $\D=0$,
\be
(r_+^2+a^2)(1+\l r_+^2)-2mr_+^2=0\,,                                 \lab{3.2}
\ee
and the angular velocity is given by
\be
\om:=\frac{g_{t\vphi}}{g_{\vphi\vphi}}
    =\frac{a(r^2+a^2-\D)\a}{(r^2+a^2)^2-a^2\D}\,,\qquad
\om_+:=\om(r_+)=\frac{a\a}{r_+^2+a^2}\,.
\ee
For large $r$, the angular velocity does not vanish, $\om=-\l a+O_2$.

The metric has two Killing vectors, $\pd_t$ and $\pd_\vphi$.
Their combination $\xi=\pd_t-\om_+\pd_\phi$ is a null vector on the horizon and normal to it. The surface gravity $\k$ can be defined by the relation $\pd_\m\xi^2=-2\k\xi_\m$ on horizon.
After rewriting the metric in terms of the ADM lapse and shift variables $n$ and $n_\vphi$, respectively:
\bsubeq\lab{3.4}
\bea
&&ds^2=n^2dt^2+g_{\vphi\vphi}(d\vphi+n_\vphi dt)^2-\frac{dr^2}{H^2}\,,   \\
&&n^2:=g_{tt}-\frac{g^2_{t\vphi}}{g_{\vphi\vphi}}=\frac{r^2}{\S^2}\D\,,
  \qquad n_\vphi:=\frac{g_{t\vphi}}{g_{\vphi\vphi}}\, ,\qquad
         H^2=\frac{\D}{r^2}\,.
\eea
\esubeq
the surface gravity can be calculated as
\be
\k=n\pd_r H\Big|_{r_+}=\frac{[\pd_r\D]_{r_+}}{2(r_+^2+a^2)}
    =\frac{r_+(\l r_+^2-a^2/r_+^2)}{(r_+^2+a^2)}\,.                 \lab{3.5}
\ee

The line element \eq{3.1} suggest the following choice for the orthonormal coframe:
\be\lab{3.6}
b^0=H\Big(dt+\frac{a}{\a}d\vphi\Big)\,,\qquad b^1=\frac{dr}{H}\,,\qquad
  b^2=\frac{1}{r}\Big(adt+\frac{r^2+a^2}{\a}d\vphi\Big)\,.
\ee
Then, horizon area is given by
\be
A=\int_H b^2=2\pi \frac{r_+^2+a^2}{\a r_+}\,,                       \lab{3.7}
\ee
the Riemannian connection reads
\be
\om^{01}=-H' b^0-\frac{a}{r^2}b^2\, ,\qquad \om^{02}=-\frac{a}{r^2}b^1\,,
\qquad \om^{12}=-\frac{a}{r^2}b^0+\frac{H}{r}b^2\, ,                \lab{3.8}
\ee
and for $\l>0$, the curvature 2-form $R^{ij}=\l b^ib^j$ is of the AdS type.

\prg{Part 2.} Our goal is to find conserved charges and entropy of the Kerr-AdS black hole \eq{3.1}, with respect to the background defined by $m=0$. However, since the background metric in Boyer-Lindquist coordinates depends on the parameter $a$, it is difficult to differentiate the background configuration from the black hole solution. Analyzing the corresponding 4D case,  Henneaux and Teitelboim \cite{x9} found that the problem can be overcome by going over to a new coordinate system, in which the asymptotic behaviour of the Kerr-AdS solution is \emph{manifestly AdS}, see also Hecht and Nester \cite{x10}. Analyzing the same problem in 3D, Hawking et al. \cite{x3} used an analogous procedure based on the coordinate transformations
\bsubeq\lab{3.9}
\bea
&& T=t\, ,\quad \phi=\vphi-\l a t\,,                              \lab{3.9a}\\
&&\r^2=\frac{r^2+a^2}{\a}+\frac{2ma^2}{\a^2}\, .                  \lab{3.9b}
\eea
\esubeq
In these coordinates, the background metric takes the standard AdS form, see section \ref{sec5}.

However, it turns out that our variational approach \eq{2.3} is well-defined even with the restricted form of the above transformation, where the radial coordinate $r$ is \emph{left unchanged}. To show that, we focus our attention on the $(T,\phi)$ transformations \eq{3.9a}. Let us first calculate the new metric components:
\bea
&&g_{TT}=g_{tt}+2\l a g_{t\vphi}+(\l a)^2g_{\vphi\vphi}
        =1+\frac{\l}{\a}(r^2+a^2)-\frac{2m}{\a^2}\,,                 \nn\\
&&g_{T\phi}=g_{t\vphi}+\l a g_{\vphi\vphi}=-\frac{2ma}{\a^2}\,,      \nn\\
&&g_{\phi\phi}=g_{\vphi\vphi}
              =-\Big(\frac{r^2+a^2}{\a}+\frac{2ma^2}{\a^2}\Big)\,,\lab{3.10}
\eea
whereas $g_{rr}$ remains unchanged, $g_{rr}=1/H^2$.
The horizon is determined by the same equation \eq{3.2}, but transition to the $(T,\phi)$ coordinates \emph{modifies} the angular velocity:
\be
\Om=\frac{g_{T\phi}}{g_{\phi\phi}}=\om+\l a\,,\qquad
\Om_+=\om_++\l a=\frac{a(1+\l r_+^2)}{r_+^2+a^2}\,.
\ee
Asymptotically, $\Om=O_2$. The value of the surface gravity does not change, as follows from formula \eq{3.5} and the relation that defines the new lapse variable $N$:
\be
N^2:=g_{TT}-\frac{g^2_{T\phi}}{g_{\phi\phi}}
           =g_{tt}-\frac{g^2_{t\vphi}}{g_{\vphi\vphi}}\equiv n^2\,.\nn
\ee

In what follows, we will use the coframe \eq{3.6} expressed in the new coordinates as
\be
b^0=\frac{H}{\a}(dT+ad\phi)\,,\qquad b^1=\frac{dr}{H}\, ,\qquad
b^2=\frac{1}{\a r}\Big[(r^2+a^2)d\phi+a(1+\l r^2)dT\Big]\, ,       \lab{3.12}
\ee
The related Riemannian connection formally coincides with the old result \eq{3.8}:
\be
\om^{01}=-H' b^0-\frac{a}{r^2}b^2\, ,\qquad \om^{02}=-\frac{a}{r^2}b^1\,,
\qquad \om^{12}=-\frac{a}{r^2}b^0+\frac{H}{r}b^2\,.
\ee

\section{Conserved charges and entropy}
\setcounter{equation}{0}

Following Henneaux and Teitelboim \cite{x9}, one could conclude that the coframe \eq{3.12}, obtained by the $(T,\phi)$ coordinate transformations \eq{3.9a}, is still not a good choice for further calculations, as its background form, defined by $m=0$, depends on the parameter $a$. However, such a conclusion needs to be reconsidered in the context of our variational approach. Namely, the rule (a1) from section 2 requires to avoid the variations $\d a$ that correspond to the background configuration. How can one recognise and eliminate these redundant $\d a$ terms? It turns out that there is a simple answer to this question:
\bitem
\item[(a1$^\prime$)] First apply $\d$ to all $a$'s, then disregard  those $\d a$'s that survive the limit $m=0$, as they stem from the unneeded variation of the background configuration with $m=0$.
\eitem
It is interesting to note that the application of this rule to the coframe \eq{3.6} in Boyer-Lindquist coordinates does not work. Indeed, the variational equations \eq{2.3} imply
\be
\d E_\text{BL}=\d\G_\infty(\pd_t)=\a\d\Big(\frac{m}{4\a^2}\Big)\, ,\qquad
\d J_\text{BL}=\d\Big(\frac{ma}{2\a^2}\Big)\, ,                   \lab{4.1}
\ee
where the expression $\d E_\text{BL}$ is \emph{not integrable}. Hence, `energy' $E_\text{BL}$ is \emph{not a well-defined} object, as opposed to the result $E_\text{BL}=m/4\a$ found in \cite{x3}. However, as we shall see, transition to the $(T,\phi)$ coordinate system \eq{3.9a} will be sufficient to consistently resolve the problem.

To continue with calculations of conserved charges based on the coframe \eq{3.12}, we display here two useful formulas:
\be
H\d H=\frac{1}{2}\d H^2=\l a\d a-\d m+O_2\,,\qquad
  HH'=\frac{1}{2}(H^2)'=\l r-\frac{a^2}{r^3}\, .                   \nn
\ee

Energy is determined by the expression $\d\G_\infty(\pd_T)$, defined in \eq{2.3}. Relying on the above rule $(a1{}^\prime)$, one finds the following nonvanishing contributions:
\bea
&&\om^{12}{_T}\d b^0
  =\frac{\l aH}{\a}\d\left(\frac{aH}{\a}\right)d\phi
  =-\d\Big(\frac{m\l a^2}{\a^2}\Big)d\phi\,,                            \nn\\
&&\d\om^{12}b^0{_T}=\frac{H}{\a}\d\left(\frac{H}{\a}\right)d\phi
  =-\d\Big(\frac{m}{\a^2}\Big)d\phi\,.
\eea
Summing up and integrating, one obtains
\bea
&&\d E=\d\G_\infty(\pd_T)
      =4\pi a_0\d\Big[\frac m{\a^2}(1+\l a^2)\Big]\, ,                 \nn\\
\Ra&&E=\frac{m}{4\a^2}(1+\l a^2)-\frac{1}{8}\,,                     \lab{4.3}
\eea
where we used $16\pi a_0=1$. The value of energy for $m=0$ is chosen to be $-1/8$, which corresponds to the AdS background.

Similarly, the angular momentum is determined by the expression $\d\G_\infty(\pd_\phi)$. Since the only nontrivial contribution is
\be
\om^{12}{}_\phi\d b^0+\d\om^{12}b^0{}_\phi
  =\d\Big(\frac{N^2a}{\a^2}\Big)d\phi
                     =\d\Big(\frac{-2ma}{\a^2}\Big)d\phi\,,
\ee
it follows
\be
\d J=\d\G_\infty(\pd_\phi)=4\pi a_0\,\d\Big(\frac{2ma}{\a^2}\Big)
\quad\Ra\quad J=\frac{ma}{2\a^2}\,.                                 \lab{4.5}
\ee

To calculate entropy, we use $\xi=\pd_T-\Om_+\pd_\phi$ in $\d\G_H(\xi)$. Starting with
\bea
&&(\xi\inn b^0)|_{r=r_+}=0\,,\qquad (\xi\inn b^2)|_{r=r_+}=0\,,       \nn\\
&&\xi\inn\om^{01}=-(1-a\Om_+)\frac{1}{\a}HH'\Big|_{r=r_+}
  =-\frac{\pd_r \D}{2(r_+^2+a^2)}\Big|_{r=r_+}\equiv -\k\, ,
\eea
one obtains
\bea
&&\d\G_H(\xi)=-\int_H(\xi\inn\om^{01})\d b^2=2a_0\k\d A\, ,             \nn\\
\Ra&&\d\G_H(\xi)=TdS\,,\qquad S=\frac{A}{4}\,,
\eea
where we used $T=\k/2\pi$.

A straightforward calculation confirms the validity of the first law:
\be
T\d S=\d E-\Om_+\d J\,.                                            \lab{4.8}
\ee

\section{Kerr-AdS metric in the BTZ form}\label{sec5}
\setcounter{equation}{0}

Here, we use an independent procedure to verify the consistency of the results found in the previous section. In the coordinate system defined by \eq{3.9}, which includes also the new radial coordinate $\r$, the Kerr-AdS metric is given by
\be
ds^2=(-8\m+\l\r^2)dT^2 -8j\,dTd\phi-\r^2 d\phi^2\, ,               \lab{5.1}
\ee
where the expressions
\be
\m=\frac{m}{4\a^2}(1+\l a^2)-\frac{1}{8}\,,\qquad j=\frac{ma}{2\a^2}\,.
\ee
coincide with the conserved charges \eq{4.3} and \eq{4.5}, respectively. Using the lapse and shift variables, the metric \eq{5.1} can be written in the BTZ (Ba\~nados-Teitelboim-Zanelli) form
\be
ds^2=N^2 dT^2-\r^2(d\phi+N_\phi dT)^2-\frac{d\r^2}{N^2}\, ,        \lab{5.3}
\ee
where
\bea
&&N^2:=g_{TT}-\frac{g^2_{T\phi}}{g_{\phi\phi}}
     =-8\mu+\l\r^2+\frac{16j^2}{\r^2}\, ,                             \nn\\
&&N_\phi:=\frac{g_{T\phi}}{g_{\phi\phi}}=\frac{4j}{\r^2}\, .       \lab{5.4}
\eea
On the AdS background with $m=0$, the integration constants are $\mu=-1/8$ and $j=0$.

Horizon is defined by the larger root of $N^2=0$:
\be
\r^2_\pm=\frac{4}{\l}\left(\m\pm\sqrt{\m^2-\l j^2}\,\,\right)\,.    \lab{5.5}
\ee
Horizon area, angular velocity and surface gravity are given by
\bea
&&A(\r_+)=2\pi\r_+\, ,\qquad
  \Om_+=\frac{4j}{\r_+^2}\,,                                          \nn\\
&&\k(\r)=\frac{1}{2}\pd_\r N^2\Big|_{\r_+}=\l\r_+-\frac{16j^2}{\r_+^3}\,.
\eea

By choosing a new coframe
\be
\hb^0=N dT\, ,\qquad \hb^1=\frac{d\r}{N}\, ,\qquad
                    \hb^2=\r(d\phi+N_\phi dT)\, ,
\ee
one can calculate the Riemannian connection as
\be
\hom^{01}=-N'\hb^0-N_\phi\hb^2\,,\qquad
\hom^{02}=-N_\phi\hb^1\, ,\qquad
\hom^{12}=-N_\phi\hb^0+\frac{N}{\r}\hb^2\,.
\ee
The basis $\hb^i$ has a particularly clean asymptotic form, in accordance with the discussion given in \cite{x9}. The variational formulas \eq{2.3} produce the results
\bea
&&\d\G_\infty(\pd_T)=-2a_0\int_{S_\infty}\d\om^{12}{}\hb^0{}_T
                       =16\pi a_0\d\m\,,                              \nn\\
&&\d\G_\infty(\pd_\phi)=-2a_0\int_{S_\infty}\d\om^{01}\hb^2{}_\phi
                       =16\pi a_0\d j\, ,
\eea
and consequently, the conserved charges take the expected form:
\be
E=\m\,,\qquad J=j\, .
\ee
Entropy is defined by the Killing vector $\xi=\pd_T-\Om_+\pd_\phi$:
\be
\d\G_H(\xi)=-2a_0\int_H(\xi\inn\om^{01})\d b^2=2a_0\k\d A
             =T\d S\,.
\ee
Finally, using the expression \eq{5.5} for $\r_+$, one can easily verify the validity of the first law in the standard form \eq{4.8}.

\section{Discussion}\label{sec6}
\setcounter{equation}{0}

In the present paper, we studied conserved charges, angular velocity and entropy as thermodynamic variables of the Kerr-AdS black hole in \gr3. These variables are found to be consistently related to each other through the first law of black hole dynamics. The result is obtained using the Hamiltonian approach to black hole entropy \cite{x6} and additionally verified by transforming the metric into the BTZ form.

One should note that our results for thermodynamic variables in Boyer-Lindquist coordinates differ from those given in Ref. \cite{x3}, section 3. To  explain the difference, consider the value of the regularized Euclidean action, analytically extended to the Minkowski region:
\be
I_3=\b F\, ,\qquad F:=\frac{m}{4\a}-\frac{\l(r^2+a^2)}{4\a}\, ,
\ee
where $\b$ is the inverse temperature. The expression for free energy $F$ can be represented in two coordinate systems, $(t,\vphi)$ and $(T,\phi)$, as follows:
\be
F(t,\vphi)=\frac{m}{4\a}-\om_+J-TS\, ,\qquad F(T,\phi)=\mu-\Om_+ J-TS\, .
\ee
In the first option, $E_\text{BL}:=m/4\a$ is interpreted as energy and $\om_+$ as angular velocity in the Boyer-Lindquist coordinates, compare with Eq. (3.22) in \cite{x3}. However, this option is not consistent, as follows from the following arguments:
\bitem\vsm
\item[i)]  the notion of energy in the Boyer-Lindquist coordinates $(t,\vphi)$ is not well defined since the variation $\d\G_\infty(\pd_t)$ is not integrable, see Eq. \eq{4.1}; \vsm
\item[ii)] the choice $E_\text{BL}=m/4\a$ does not comply with the first law.
\eitem\vsm
On the other hand, the second option offers a consistent interpretation of $\m$ as energy and $\Om_+$ as angular velocity of the Kerr-AdS black hole, in complete agreement with the standard form \eq{4.8} of the first law.

We plan to extend the present approach to the corresponding 4D problem \cite{x2}.



\begin{thebibliography}{99}

\bibitem{x1} R. M. Wald, Black hole entropy is the Noether charge, Phys. Rev. D {\bf 48} (1993) R3427-R3431;
        	
    T. Jacobson, G. Kang, and R. C. Myers, On black hole entropy,
    Phys. Rev. D {\bf 49} (1994) 6587-6598.

\bibitem{x2} G. W. Gibbons, M. J. Perry, and C. N. Pope, The first law of thermodynamics for Kerr-anti-de Sitter black holea, Class. Quant. Grav. {\bf 22} (2005) 1503-1526.

\bibitem{x3} S. W. Hawking, C. J. Hunter, and M. M. Taylor-Robinson, Rotation and the AdS/CFT correspondence, Phys. Rev. D {\bf 59} (1999) 064005 (13 pages).

\bibitem{x4} H. Kim, Spinning BTZ black hole versus Kerr black hole: a closer look, Phys. Rev. D {\bf 59} (1999) 064002 (7 pages).

\bibitem{x5} W. Shuang, W. Shuang-Qing, X. Fei, and D. Lin, 	
    The First laws of thermodynamics of the (2+1)-dimensional BTZ black holes and Kerr-de Sitter spacetimes, Chin. Phys. Lett. {\bf 23} (2006) 1096-1098.

\bibitem{x6} M. Blagojevi\'c and B. Cvetkovi\'c, Entropy in Poincar\'e gauge theory: Hamiltonian approach, Phys. Rev. D 99 (2019) 104058 (12 pages).

\bibitem{x7} M. Blagojevi\'c and B. Cvetkovi\'c, Hamiltonian approach to black hole entropy: Kerr-like spacetimes,  Phys. Rev. D {\bf 100} (2019) 044029 (7 pages).

\bibitem{x8} M. Blagojevi\'c and F. W. Hehl (eds.), \emph{Gauge Theories of Gravitation, A Reader with Commentaries} (Imperial College Press, London, 2013); arXiv:1210.3775 [gr-qc].

\bibitem{x9} M. Henneaux and C. Teitelboim, Asymptotically anti-de Sitter spaces, Commun. Math. Phys. {\bf 98}  (1985) 391-424.

\bibitem{x10} R. D. Hecht and J. M. Nester, A new evaluation of PGT mass and spin, Phys. Lett. A {\bf 180} (1993) 324-331.

\end{thebibliography}
\end{document}